\documentclass[aps,pra,twocolumn, showpacs]{revtex4}
\usepackage{amsmath, amsthm, amssymb,amsbsy,graphicx,times,subfigure,marvosym,color,bm,hyperref}
\usepackage[latin1]{inputenc}
\usepackage{sidecap}

\usepackage{ulem}

\begin{document}

\title{Quantum Fluctuations of Vortex Lattices in Ultracold Gases}
\author{M. P. Kwasigroch}
\author{N. R. Cooper}
\address{T.C.M. Group, Cavendish Laboratory, University of Cambridge, J. J. Thomson Avenue, Cambridge CB3 0HE, UK }


\begin{abstract}

  We discuss the effects of quantum fluctuations on the
    properties of vortex lattices in rapidly rotating ultracold atomic
    gases.  We develop a variational method that goes beyond the
    Bogoliubov theory by including the effects of interactions between
    the quasiparticle excitations.  These interactions are found to
    have significant quantitative effects on physical properties even at relatively large filling factors. We use our theory to predict the expected experimental signatures of
    quantum fluctuations of vortices, and to
    assess the competition of the triangular vortex lattice phase with
    other phases in finite-sized systems.

\end{abstract}

\maketitle


\section{Introduction}\label{intro}

 The achievement of Bose-Einstein condensation in dilute atomic
  gases has allowed studies of the interesting collective
  behaviour of these quantum fluids with unprecedented detail and in
  previously inaccessible parameter regimes~\cite{blochdz}. The interatomic
  interactions, although weak, cause the atomic Bose-Einstein
  condensates to exhibit superfluidity and the associated phenomena
  familiar from studies of superfluid helium. Among the most striking
  of these features is the appearance of quantized vortices when the
  superfluid is forced to rotate~\cite{Donnelly, fetterreview}. Experiments on rapidly rotating
  atomic gases have shown beautiful images of vortex
  lattices~\cite{madison,aboshaeer,SchweikhardCEMC92}, and have allowed detailed studies
  of their collective properties~\cite{cornell1,foot}.

  Theoretical studies have pointed to the possibility of using
  ultracold atomic gases to explore novel regimes of vortex density
  where quantum fluctuations of the vortices become
  important~\cite{CooperReview}.  The parameter controlling the degree
  of quantum fluctuations is the filling factor, $\nu\equiv N/N_{\rm
    v}$, where $N$ and $N_{\rm v}$ denote the number of bosons and
  vortices respectively~\cite{QuantumCooper}. Quantum fluctuations are
  small for $\nu\to \infty$, but increase with decreasing $\nu$. In
  particular, for a large uniform system, it has been shown that these
  fluctuations drive a phase transition out of the triangular vortex
  lattice to a set of strongly correlated bosonic quantum liquids (analogous
  to fractional quantum Hall states) when the filling factor is
  reduced below a critical value $\nu_{\rm
    c}$~\cite{QuantumCooper}. Estimates of $\nu_{\rm c}$ vary between
  $\nu_{\rm c}\simeq 2-6$ from exact diagonalization 
  results~\cite{QuantumCooper,CooperSize,vedral} and $\nu_{\rm c}\simeq
  8-14$ from a Lindemann criterion~\cite{QuantumCooper,Sinova,Baym}.

  While the prediction of the precise transition point is a very
  difficult theoretical task, the analytical theories of
  Refs.~\cite{Sinova,Baym,Sonin,Matveenko} do also provide important
  information concerning the effects of quantum fluctuations within the
  vortex lattice  phase. These
  results are based on the application of mean-field theory, with
  quantum fluctuations described by the Bogoliubov approximation. This
  is expected to be accurate in the limit of large filling factors,
  $\nu \gg 1$~\cite{QuantumCooper}.  However, for smaller values of $\nu$, one expects
  corrections to the results of Bogoliubov theory, arising from the
  effects of quartic terms in the fluctuation expansion, or
  equivalently from effective interactions between the Bogoliubov
  quasiparticles. In this paper, we analyse the effects of these
  corrections to the Bogoliubov theory. We introduce a variational
  wavefunction for the vortex lattice of a rapidly rotating atomic
  Bose gas, which recovers the Bogoliubov theory at $\nu\to \infty$
  but which incorporates the effects of interactions between the
  Bogoliubov quasiparticles. We show that the variational theory
  accurately reproduces the exact diagonalization results for $\nu\gtrsim 5$ in the small
  systems for which these results are available.  We then apply the
  theory to large systems, representative of the thermodynamic limit.
  We find that the quasiparticle interactions lead to large
  quantitative corrections to the predictions of Bogoliubov theory even for large filling
  factors deep in the
  vortex lattice regime. We discuss how effects of quantum
  fluctuations may be observed in experiments measuring the condensate
  fraction, particle density at the vortex core or the shear modulus of the vortex lattice. Finally, we
  discuss the implications of our results for the competition with other
  phases on finite-sized systems.

  The paper is organised as follows: In Sec.~\ref{model} we review
  the basic model for our system and derive the form of the
  microscopic Hamiltonian in the rapid-rotation limit. We then apply
  Bogoliubov theory in Sec.~\ref{Bgbv}, obtaining the spectrum of
  low-energy excitations for the triangular lattice, and using this to
  derive the effects of quantum fluctuations on various physical observables. In
  Sec.~\ref{variational}, we present a variational approach which
  allows studies beyond the Bogoliubov limit of very small condensate
  depletion.  In Sec~\ref{ED}, we compare the results of this
  variational theory with small system exact diagonalization results, and describe its predictions both for (thermodynamically) large systems and for
  finite-sized systems. Sec~\ref{Summary} contains a summary of our results.

\section{Theoretical Model} \label{model}

A system of $N$ interacting bosons of mass $M$ in a parabolic trap that is rotating with frequency $\Omega$ about the $z$-axis can be described by the following rotating-frame Hamiltonian~\cite{CooperReview}:
\begin{eqnarray}
H &=& \displaystyle\sum_{i=1}^{N}  \Big{\{} \frac{(\mathbf{p}_i - M\Omega\hat{\mathbf{z}}\times \mathbf{r}_i)^2}{2 M} \nonumber\\
&&+\frac{M}{2}\big{[}(\Omega_r^2 - \Omega^2)(x_i^2+y_i^2)+\Omega_z^2z_i^2\big{]} \Big{\}} \nonumber \\
&&+ \displaystyle\sum_{i<j=1}^{N}V(\mathbf{r}_i - \mathbf{r}_j),\label{Hamiltonian}
\end{eqnarray}
where $\Omega_r$ and $\Omega_z$ are the radial and axial trap frequencies respectively and $V(\mathbf{r}) = g_{\rm{3D}}\delta(\mathbf{r})$ is a contact interaction of strength $g_{\rm{3D}}$. The Hamiltonian is equivalent to a system of charge-$q$ bosons in an effective magnetic field $\mathbf{B}=(2M\Omega/q)\hat{\mathbf{z}}$. We will be working under the assumptions of weak interactions and quasi 2-dimensionality where the inequalities $ng\ll\hbar\Omega$ and $ng\ll\hbar\Omega_z$  are satisfied ($n$ is the 2D boson density and $g=g_{\rm{3D}}\sqrt{\frac{M\Omega_z}{2 \pi \hbar}}$ is the 2D contact interaction parameter)~\cite{AxialConfinement}. We set $\Omega_r=\Omega$ so that any residual radial confinement can be neglected. Treating the interaction term as a perturbation, it follows that the energy eigenstates of $H$ are Landau levels~\cite{CooperReview}, with eigenvalues given by $E_n=2\hbar\Omega(n+1/2)$, and that all the particles are confined to the lowest Landau level (LLL). We shall refer to the above theoretical set-up as the 2D LLL regime. 
The LLL basis states take the following form:
\begin{equation}
\Phi_j(\mathbf{r})=\frac{1}{\sqrt{2\pi l^2 2^j j!}} (x+iy)^j e^{-(x^2+y^2)/4l^2},
\end{equation}
where $l=\sqrt{\hbar/2M\Omega}$ is the effective magnetic length and $j=0, 1, 2, \ldots$ indexes the states. 

Within mean-field theory, the condensate wavefunction, $\Psi(\mathbf{r})$, is a superposition of these basis states and is determined solely by the positions of its zeros~\cite{bournewilkingunn}. These correspond to coordinates of vortex centres. Under the requirement that the entire system mimics rigid body motion the number of vortices in the system must be given by
\begin{equation}
N_{\rm v}=A/(2\pi l^2), \label{vortex}
\end{equation}
where $A$ is the condensate area.

Studying the structure of the mean-field ground state will help us to choose an appropriate basis in which to treat quantum fluctuations. In the 2D LLL regime we are left to consider only the interaction term in the Hamiltonian. In the mean-field approximation the interaction term reduces to
\begin{equation}
\frac{gN^2}{2}\int|\Psi(\mathbf{r})|^4\,d^2\mathbf{r},
\end{equation}
 and assumes a minimum when the zeros of $\Psi(\mathbf{r})$ lie on a triangular lattice~\cite{CooperReview}. Thus, the mean-field ground state is a triangular lattice of vortices with a unit cell area of $2\pi l^2$ (a consequence of Eq.~\ref{vortex}).

As our basis, we will use LLL magnetic Bloch states which are related to the mean-field ground state by a translation. We review their construction, outlined in Ref.~\cite{Rashba}. One starts with the most localized symmetric LLL wavefunction
\begin{equation}
c_0(\mathbf{r})=\Phi_{j=0}(\mathbf{r})=\frac{1}{\sqrt{2\pi l^2}}e^{-\mathbf{r}^2/4l^2},
\end{equation}
and translates it to the sites of a 2D Bravais lattice, to obtain
\begin{eqnarray}
c_\mathbf{m}(\mathbf{r})&=&T_{m_1 \mathbf{a}_1}T_{m_2 \nonumber \mathbf{a}_2}c_0(\mathbf{r})\\&=&\frac{(-1)^{m_1m_2}}{\sqrt{2\pi l^2}}e^{-(\mathbf{r}-\mathbf{r_m})^2/4l^2+(i/2l^2)\hat{\mathbf{z}}\cdot(\mathbf{r}\times \mathbf{r_m})},\nonumber\\
\end{eqnarray}
where $\mathbf{a}_1$ and $\mathbf{a}_2$ are the Bravais lattice vectors, $\mathbf{r_m}=m_1\mathbf{a}_1+m_2\mathbf{a}_2$ is the translation vector from the origin to a given lattice site, and $T_\mathbf{\mathbf{r_m}}=\exp\left[-\frac{i}{\hbar}\mathbf{r_m}\cdot(\mathbf{p}+m\Omega\hat{\mathbf{z}}\times \mathbf{r})\right]$ is the corresponding translation operator~\cite{Zak}. From Eq.~\ref{vortex}, it follows that one quantum flux passes through the unit cell area $|\mathbf{a}_1\times\mathbf{a}_2|$, and therefore, the magnetic translation operators commute with the Hamiltonian and each other, generating the elements of the Magnetic Translation Group~\cite{Zak}. Following Ref.~\cite{Rashba}, we can now construct orthonormal Bloch functions out of linear combinations of $c_{\mathbf{m}}(\mathbf{r})$
\begin{equation}
\Psi_\mathbf{k}(\mathbf{r})=\frac{1}{\sqrt{N_{\rm v}\zeta(\mathbf{k})}}\sum_\mathbf{m}c_\mathbf{m}(\mathbf{r})e^{i\mathbf{k}\cdot\mathbf{r_m}},
\end{equation}
where $\zeta(\mathbf{k})$ normalises the Bloch function to unity over the system area, and is given by
\begin{equation}
\zeta(\mathbf{k})=\sum_\mathbf{m}(-1)^{m_1m_2}e^{-\mathbf{r_m}^2/4l^2}e^{-i\mathbf{k\cdot r_m}}.
\end{equation}
The Bloch states are simultaneous eigenstates of the Hamiltonian and the translation operators. In a finite system of area $A$, there are $N_{\rm v}$ (Eq.~\ref{vortex}) such states, labelled by momenta $\mathbf{k}$, which belong to the first Brillouin zone (BZ) of the 2D Bravais lattice and satisfy periodic boundary conditions with respect to magnetic translations. We shall study a system of vortices defined on a rectangular region with dimensions $L_x$, $L_y$, and with periodic boundary conditions.

In the magnetic Bloch basis the contact interaction takes the following form:
\begin{equation}
H_{\rm I}=\sum_{\mathbf{k},\mathbf{q},\mathbf{q'},\mathbf{q'', G}}M(\mathbf{k},\mathbf{q},\mathbf{q'},\mathbf{q''})
\delta_{\mathbf{q}+\mathbf{q'},\mathbf{q''}+\mathbf{G}}
b^{\dagger}_{\mathbf{k}+\mathbf{q}}b^{\dagger}_{\mathbf{k}+\mathbf{q'}}b_{\mathbf{k}+\mathbf{q''}}b_{\mathbf{k}}, \label{interaction}
\end{equation}
where all momenta belong to the first BZ and $\mathbf{G}$ are the reciprocal lattice vectors. The Kronecker delta imposes conservation of momentum modulo a reciprocal lattice vector and the interaction matrix element, obtained in Ref.~\cite{Burkov}, is given by
\begin{eqnarray}
M(\mathbf{k},\mathbf{q},\mathbf{q'},\mathbf{q''})&=&\frac{g}{2}\int\,d^2\mathbf{r}
\Psi^\ast_{\mathbf{k}+\mathbf{q}}(\mathbf{r})\Psi^\ast_{\mathbf{k}+\mathbf{q'}}(\mathbf{r})
\Psi_{\mathbf{k}+\mathbf{q''}}(\mathbf{r})\Psi_{\mathbf{k}}(\mathbf{r})\nonumber \\
&=&\frac{g/(4A)}{\sqrt{\zeta(\mathbf{k}+\mathbf{q})\zeta(\mathbf{k}+\mathbf{q'})\zeta(\mathbf{k}+\mathbf{q''})
\zeta(\mathbf{k})}} \nonumber \\
&&\times \sum_{\mathbf{m_1}, \mathbf{m_2}, \mathbf{m_3}}(-1)^{m_{11}m_{12}+m_{21}m_{22}+m_{31}m_{32}} \nonumber \\
&&\times e^{-\frac{1}{8l^2}\mathbf{[r_{m_1}^2+r_{m_2}^2+(r_{m_1}-r_{m_3})^2+(r_{m_2}-r_{m_3})^2]}} \nonumber \\
&&\times e^{\frac{i}{4l^2}\mathbf{\hat{z}\cdot[(r_{m_1}+r_{m_2})\times r_{m_3}]}-i(\mathbf{k+q)\cdot r_{m_1}}} \nonumber \\
&&\times e^{-i\mathbf{(k+q')\cdot r_{m_2}}+i\mathbf{(k+q'')\cdot r_{m_3}}}, \nonumber\\ \label{matrix}
\end{eqnarray}

 The mean-field ground state corresponds to all $N$ particles condensed into the $\mathbf{k}=\mathbf{0}$ Bloch state. It has energy $E=N^2M(\mathbf{0,0,0,0})=\beta_{\triangle}\frac{gN^2}{2A}$, where $\beta_{\triangle}\simeq 1.1596$.

One can readily extend this approach to calculate the shear modulus within mean-field theory. The energy change per unit area of a vortex lattice, sheared by an infinitesimal angle $\theta$, is given by $C_2\theta^2$, where $C_2$ is its shear modulus~\cite{Sonin}. The shear modulus can be calculated in the mean-field approximation as 
\begin{equation}
C_2=\frac{N^2}{A}\partial^2M(\mathbf{0,0,0,0})/\partial\theta^2=0.1191g n^2,
\end{equation}
where the $\theta$ dependence of $M(\mathbf{0,0,0,0})$ enters through the sheared lattice vectors:
\begin{eqnarray}
&&\mathbf{a}_1=a_{\triangle}(1,0) \nonumber \\
&&\mathbf{a}_2=\frac{a_{\triangle}}{2}(1+\theta\sqrt{3},\sqrt{3}). \label{ShearedVectors}
\end{eqnarray}
$a_{\triangle}=(\frac{4\pi}{\sqrt{3}})^{\frac{1}{2}}l$ is the lattice constant of the triangular vortex lattice.
The shear modulus is relevant for the distortion of the lattice
  by thermal fluctuations, and ultimately the thermal melting
  transition. (Though non-linearities from thermal fluctuations will
  become important.) It also determines the frequencies of the
  long-wavelength sound modes in the hydrodynamic regime. This limit
  applies when the scattering rate $\Gamma_{\mathbf k}$ of Bogoliubov
  modes is large compared to their energy $\epsilon_{\mathbf k}$.
  Matveenko et al.~\cite{Matveenko} have shown that at zero
  temperature,
  $\hbar\Gamma(\epsilon(\mathbf{k}),0)/\epsilon(\mathbf{k})\simeq
  0.065/\nu$ so the system is always lightly damped at filling factors
  where the triangular lattice is predicted to exist
  (i.e. ${\nu\gtrsim 7}$). However, at non-zero temperature,
excitations with  $\epsilon(\mathbf{k})\lesssim \epsilon_c=\frac{T}{\nu}$ are overdamped (see  ~\cite{Matveenko}). Thus, for small non-zero temperatures, one can 
 expect these modes to be in the hydrodynamic regime, and have frequencies set by the shear modulus.

It is useful to note that the zeros of the Bloch function $\Psi_{\mathbf{k}}(\mathbf{r})$ are located at~\cite{Burkov}
\begin{equation}
\mathbf{r_m}+\frac{1}{2}(\mathbf{a}_1+\mathbf{a}_2)+l^2\mathbf{\hat{z}\times k}.
\end{equation}
From this, we see that the zeros map out a triangular lattice which is simply a translation of the mean-field $\mathbf{k}=\mathbf{0}$ lattice. Adding quantum fluctuations to the mean-field theory leads to excitations of neighbouring $\mathbf{k}$-states, which correspond to vortex lattices that are offset with respect to the macroscopically occupied one. We can say that quantum fluctuations ``smear'' the original mean-field lattice.

\section{Bogoliubov Approximation}  \label{Bgbv}

At high filling factors, the condensate is weakly depleted, and we can approximate the Hamiltonian by a Bogoliubov expansion, which keeps terms up to second order in $(N-N_0)$, where $N_0$ is the macroscopic occupation number of the condensate. The condensate is a coherent state, i.e. an eigenstate of $b_0$. Its creation and annihilation operators can be treated as $c$ numbers given by
\begin{equation}
 b^{\dagger}_{0}= b_{0}=\sqrt{N_0}=\sqrt{N-\sum_{\mathbf{q}\neq\mathbf{0}}\langle b^{\dagger}_{\mathbf{q}}b_{\mathbf{q}}\rangle}.
\end{equation}
 The Bogoliubov expansion of $H_{\rm I}$ takes the following explicit form:
\begin{eqnarray}
H_{\rm I}&=& M(\mathbf{0,0,0,0})N^2 \nonumber \\
&&+ \frac{N}{2}\sum_{\mathbf{q}\neq\mathbf{0}} \left(b^{\dagger}_{\mathbf{q}}, b_{-\mathbf{q}}\right)\left(\begin{array}{cc}
\xi_\mathbf{q} & |\lambda_\mathbf{q}|e^{i\phi_\mathbf{q}} \\
|\lambda_\mathbf{q}|e^{-i\phi_\mathbf{q}} & \xi_\mathbf{q}
\end{array}\right)\left(\begin{array}{c}b_\mathbf{q}\\b^{\dagger}_{-\mathbf{q}}\end{array}\right)\nonumber \\
&&-\frac{N}{2}\sum_{\mathbf{q}\neq\mathbf{0}}\xi_\mathbf{q} + \mathcal{O}\left((N-N_0)^2\right),
\end{eqnarray}
where
\begin{eqnarray}
&&|\lambda_\mathbf{q}|e^{i\phi_\mathbf{q}}= 2M(0,\mathbf{q},-\mathbf{q},0), \nonumber\\
&&|\lambda_\mathbf{q}|e^{-i\phi_\mathbf{q}}=2M(\mathbf{q},-\mathbf{q},-\mathbf{q},-2\mathbf{q}), \nonumber \\
&&\xi_\mathbf{q} = 2\left[M(\mathbf{0,0,q,q})+M(\mathbf{0,q,0,q})-M(\mathbf{0,0,0,0})\right].\nonumber \\
\end{eqnarray}
In this approximation the Hamiltonian is quadratic and can be diagonalized via a Bogoliubov transformation of the form:
\begin{eqnarray}
\left(\begin{array}{c}b_\mathbf{q} \\ b^{\dagger}_{-\mathbf{q}}\end{array}\right)=\left(\begin{array}{cc}
\cosh\theta_\mathbf{q} & -e^{i\phi_\mathbf{q}}\sinh\theta_\mathbf{q}  \\
-e^{-i\phi_\mathbf{q}}\sinh\theta_\mathbf{q}  & \cosh\theta_\mathbf{q}\end{array}\right)\left(\begin{array}{c}\alpha_\mathbf{q}\\\alpha^{\dagger}_{-\mathbf{q}}
\end{array}\right), \nonumber \\ \label{Bogoliubov}
\end{eqnarray} to give
\begin{eqnarray}
H_{\rm I}&=& N^2M(\mathbf{0,0,0,0})\nonumber\\
&&+N\sum_{\mathbf{q}\neq\mathbf{0}} \left[\sqrt{\xi_\mathbf{q}^2-|\lambda_\mathbf{q}|^2}(\alpha^{\dagger}_{\mathbf{q}}\alpha_\mathbf{q} + \frac{1}{2}) - \frac{1}{2}\xi_\mathbf{q}\right],\nonumber\\
\label{HI}
\end{eqnarray}
where $\tanh2\theta_\mathbf{q}=\frac{|\lambda_\mathbf{q}|}{\xi_\mathbf{q}}$. The number of particles outside the condensate is given by
\begin{eqnarray}
N-N_0&=&\sum_{\mathbf{q}\neq\mathbf{0}}\langle b^{\dagger}_{\mathbf{q}}b_\mathbf{q}\rangle = \sum_{\mathbf{q}\neq\mathbf{0}}\sinh^2\theta_\mathbf{q},  \label{deplet}
\end{eqnarray}
which is assumed small compared to $N$ for self-consistency of the mean field approach. However, in the vicinity of $\mathbf{q=0}$, one finds that $\sinh^2\theta_{\mathbf{q}}\sim 1/(ql)^2$. Using
 $\sum_\mathbf{q}
\rightarrow \frac{l^2N_{\rm v}}{2\pi}\int\,d^2\mathbf{q}$, one finds that
 the depleted fraction diverges with system size~\cite{Sinova} as $(N-N_0)/N \sim \ln(N_{\rm v})/\nu$.

The excitation spectrum of the triangular lattice, given by
  Eq.~\ref{HI}, agrees with previous
  calculations~\cite{Sinova,Matveenko}. We extend these works on
  the Bogoliubov approximation by evaluating also the zero-point energy, the particle density at the vortex core, and the
  shear modulus. To do so, we
  calculate the matrix elements, $\xi_\mathbf{q}$ and
  $|\lambda_\mathbf{q}|$, by evaluating the sums in Eq.~\ref{matrix}
  for the triangular lattice. (See the appendix for an outline of this
  calculation.) 

We find that including the zero-point fluctuations within Bogoliubov theory leads to the ground state energy
\begin{equation}
E=\frac{gN^2}{2A}\left(\beta_\triangle+\frac{\gamma_\triangle}{\nu}\right),\label{energyTr}
\end{equation}
where $\gamma_\triangle \simeq -0.5036$. 
Zero-point fluctuations give rise to a correction that increases as we move away from the mean-field regime and is inversely proportional to the filling factor.

\begin{figure}
\centering
\includegraphics[width=0.8\columnwidth, angle=270]{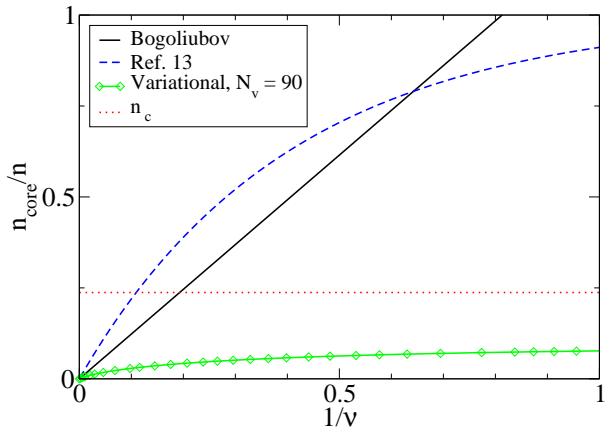}   
\caption{Vortex core particle density as a function of the inverse filling factor, for the full Bogoliubov theory (solid line), the long-wavelength theory of Ref.~\cite{Sinova} (dashed line), and for our variational wavefunction  in Eq.~\ref{definite} for $N_{\rm v}=90$ (diamonds).
 The relation given in the appendix was used to map from the mean-squared vortex core displacement $\langle\delta\mathbf{r}^2\rangle$ calculated in Ref.~\cite{Sinova} to vortex core density. $n_{\rm c}$ (dotted line) is the critical core density at which the lattice should ``melt'' according to a Lindemann criterion with  parameter $\langle\delta\mathbf{r}^2\rangle= 0.145l^2$.} \label{core}
\end{figure}

Condensate depletion, which takes place beyond mean-field, gives a non-zero expectation value of the particle density at the centre of the vortex:
\begin{equation}
n_{\rm{core}}=\sum_{\mathbf{q}\neq\mathbf{0}}\langle b^{\dagger}_{\mathbf{q}}b_\mathbf{q}\rangle |\Psi_\mathbf{q}(\mathbf{r_{\rm core}})|^2,
\end{equation}
where $\mathbf{r_{\rm core}}=\frac{1}{2}(\mathbf{a}_1+\mathbf{a}_2)$ gives the position of one of the zeros of the mean-field $\mathbf{k=0}$ wavefunction. In the Bogoliubov approximation, we find
\begin{equation}
n_{\rm{core}}/n\stackrel{N_{\rm v}\rightarrow\infty}{=}1.23/\nu.
\end{equation}

The vortex core density can be used to estimate $\nu_{\rm{c}}$, the critical filling factor at which the vortex lattice melts, by relating it to spatial fluctuations of the mean-field lattice. As was explained in the preceding section, condensate depletion leads to excitation of $\mathbf{k\neq 0}$-states, and this can be viewed as the fluctuating mean-field lattice. The fluctuations lead to a non-vanishing particle density at the core of each vortex. Assuming that the fluctuations are described by a Gaussian distribution we can write down the average vortex core density as a function of the variance of this distribution, $\langle\delta\mathbf{r}^2\rangle$, where $\delta\mathbf{r}$ is the displacement of the vortex core from its mean-field position (see the appendix for this relation). For $|\delta \mathbf{r}| \ll l$ the density is a function of the variance only and does not depend on the particular distribution of the fluctuations. One can estimate the melting point of the lattice using a suitable Lindemann criterion \cite{SinovaLind} i.e. the lattice is said to melt when $\langle\delta\mathbf{r}^2\rangle/l^2$ reaches a particular threshold value. We follow Refs.~\cite{QuantumCooper, Baym} and take this to be $\sim 0.145$.  We also calculate the vortex core density from the estimate of Ref.~\cite{Sinova} for $\langle\delta\mathbf{r}^2\rangle$, obtained through a path-integral formalism. Fig.~\ref{core} compares the core density in this long-wavelength limit with the result that we find within the Bogoliubov approximation. The Bogoliubov theory is more accurate because it uses the energy dispersion across the entire Brillouin zone, whereas the long-wavelength approximation of Ref.~\cite{Sinova} uses a quadratic approximation for the energy dispersion which neglects the contribution of higher order $\mathbf{k}$ terms to $\langle\delta\mathbf{r}^2\rangle$. The inclusion of the full dependence across the Brillouin zone leads to a change in the melting transition predicted by the Lindemann criterion with parameter $\langle\delta\mathbf{r}^2\rangle/l^2 = 0.145$ from $\nu_{\rm c} \simeq 9$ to $\nu_{\rm c} \simeq 5$. As we shall discuss below, the inclusion of effects beyond the Bogoliubov theory have an even more dramatic effect.

 The zero-point fluctuations also give a significant correction to the shear modulus,
  arising from the variation of the zero-point energy with the shear
  angle.  Under the Bogoliubov approximation we find
\begin{eqnarray}
C_2&&=\frac{1}{A}\frac{\partial^2}{\partial\theta^2}\left[N^2M(\mathbf{0,0,0,0})+\frac{N}{2}\sum_{\mathbf{q \neq 0}}\left(\sqrt{\xi_\mathbf{q}^2-|\lambda_\mathbf{q}|^2}- \xi_\mathbf{q}\right)\right]\nonumber \\
&&\stackrel{N_{\rm v}\rightarrow\infty}{=}g n^2\left(0.1191-\frac{0.1332}{\nu}\right), \label{shear}
\end{eqnarray}
where the $\theta$ dependence enters through the sheared lattice vectors (see Eq.~\ref{ShearedVectors}).

\section{Beyond The Bogoliubov Approximation}\label{variational}

The Bogoliubov expansion of the interaction Hamiltonian (Eq.~\ref{interaction}) neglects those scattering processes which do not involve condensed particles, i.e. terms of the form $b^{\dagger}_{\mathbf{k}}b^{\dagger}_{\mathbf{q}}b_{\mathbf{q'}}b_{\mathbf{q''}}$, and is only valid for small condensate depletions when $\nu\gg1$. Previous studies, summarized in Section~\ref{Bgbv}, have overlooked the contribution of this quartic term. As will be shown, the contribution of the quartic term to the energy of the system can be very significant even at large filling factors. In fact, the larger the system the higher the filling factor at which we expect the quartic term to be a significant contribution to the Bogoliubov approximation. This is a result of the diverging depletion in the vicinity of $\mathbf{k=0}$, where $\sinh^2\theta_{\mathbf{k}}\sim N_{\rm v}$ within Bogoliubov theory.
A scaling argument shows that this divergence causes the contribution of the quartic terms to the total energy to be of order $N_{\rm v}/\nu$ times the contribution of the quadratic terms. 
Hence, the Bogoliubov approximation is quantitatively valid, and the quartic term can be neglected, only when $N_{\rm v}/\nu \ll 1$.

The most straightforward way to include the quartic term in energy calculations is through a variational analysis. We take a trial wavefunction of the form obtained within the Bogoliubov approximation (retaining up to quadratic terms), parameterised by a set of variational variables. We then find the values of these parameters that optimise the full energy, including the quartic interactions. The Bogoliubov wavefunction can be expressed in terms of a depletion operator acting on a simple condensate. The operator excites and de-excites pairs of particles with a vanishing total momentum. The excitations are weighted by a set of real variational parameters, $\theta_\mathbf{k}$ and $\phi_\mathbf{k}$, indexed by the BZ momenta.
\begin{equation}
|\Psi_{\rm{trial}}\rangle=U|0\rangle, \label{trial}
\end{equation}
where $|0\rangle$ is the condensate of all particles in the ${\mathbf k}=0$ state, and $U$ a unitary depletion operator given by
\begin{equation}
U=\exp\left[-\sum_{\mathbf{k}\neq \mathbf{0}} \frac{\theta_\mathbf{k}}{2}\left(b^{\dagger}_{\mathbf{k}}b^{\dagger}_{-\mathbf{k}}e^{i\phi_\mathbf{k}}-b_{\mathbf{k}}b_{-\mathbf{k}}e^{-i\phi_\mathbf{k}}\right)\right].
\end{equation}
By making use of the following identities:
\begin{eqnarray}
U^{\dagger}b_\mathbf{q}U\equiv b_\mathbf{q}\cosh\theta_\mathbf{q} -b^{\dagger}_{-\mathbf{q}}e^{i\phi_\mathbf{q}}\sinh\theta_\mathbf{q}, \nonumber \\
U^{\dagger}b^{\dagger}_{\mathbf{q}}U\equiv b^{\dagger}_{\mathbf{q}}\cosh\theta_\mathbf{q} -b_{-\mathbf{q}} e^{-i\phi_\mathbf{q}}\sinh\theta_\mathbf{q},\label{identity}
\end{eqnarray}
 we can express $\langle \Psi_{\rm{trial}}|H_{\rm I}|\Psi_{\rm{trial}}\rangle$ as an explicit function of the parameters $\theta_\mathbf{k}$ and $\phi_\mathbf{k}$:
\begin{widetext}
\begin{eqnarray}
&&\langle \Psi_{\rm{trial}}|H_{\rm I}|\Psi_{\rm{trial}}\rangle = \sum_{\mathbf{k}\neq \mathbf{0},\mathbf{q}\neq\mathbf{0}}\Big\{2\sinh^2\theta_\mathbf{k} \sinh^2\theta_\mathbf{q}
M(\mathbf{k,0,-k+q,-k+q})\nonumber\\
&&+M(\mathbf{0,0,0,0})N_{0}^{2}+\sinh\theta_\mathbf{q}\cosh\theta_\mathbf{q}\sinh\theta_\mathbf{k}\cosh\theta_\mathbf{k} e^{i(\phi_\mathbf{k}-\phi_\mathbf{q})}M(\mathbf{k,-k+q,-q-k,-2k})\Big\}\nonumber\\ 
&&+N_0 \sum_{\mathbf{q}\neq\mathbf{0}} \Big\{4\sinh^2\theta_\mathbf{q}M(\mathbf{0,q,0,q})
-\sinh\theta_\mathbf{q}\cosh\theta_\mathbf{q} \Big[e^{i\phi_\mathbf{q}}M(\mathbf{q,-q,-q},-2\mathbf{q})+ e^{-i\phi_\mathbf{q}}M(\mathbf{0,q,-q,0})\Big]\Big\},
\end{eqnarray}
\end{widetext}
where $N_0$ can be eliminated using Eq.~\ref{deplet}. 

The Bogoliubov trial wavefunction does not have a definite particle number. The finite variance of $N$ leads to a significant overestimate of the energy at low filling factors and in small systems, i.e. whenever $N$ is not much greater than 1. We can see this from how the energy scales with $N$ in the mean-field regime: $\frac{g\beta_\triangle}{2A} N^2$. It follows that $E/N^2$ is overestimated  by $\approx\frac{g\beta_\triangle}{2 A }N^{-1}$. To address this problem we could project the Bogoliubov trial wavefunction onto a state with a definite particle number and compute the new expectation value of $H_{\rm I}$. We instead propose to use, right from the beginning, a trial wavefunction with a definite particle number of the following form, in line with work by Girardeau and Arnowitt~\cite{Bosons}:

\begin{equation}
\exp\left[-\sum_{\mathbf{k}\neq\mathbf{0}} \frac{\theta_\mathbf{k}}{2}\left(\beta b^{\dagger}_{\mathbf{k}}b^{\dagger}_{-\mathbf{k}}e^{i\phi_\mathbf{k}}-\beta^{-1}b_{\mathbf{k}}b_{-\mathbf{k}}e^{-i\phi_\mathbf{k}}\right)\right]|N\rangle,\label{definite}
\end{equation}
where $\beta^{\frac{1}{2}}=b_0(b^{\dagger}_0b_0)^{-\frac{1}{2}}$ and $|N\rangle$ is a condensate of $N$ particles in the $\mathbf{k}=\mathbf{0}$ state. The depletion operator is now particle-conserving. The expectation value of $H_{\rm I}$ for this trial wavefunction is expressed in terms of the variational parameters in the appendix. The optimal values of the parameters were found numerically for each system that we studied.

\section{Results and Comparisons with other states }\label{ED}

\subsection{Triangular Vortex Lattice}

In order to  test our variational  method we have calculated the energies of small systems and have compared them with the data available from exact diagonalisation (ED) studies of Ref.~\cite{QuantumCooper}. Variational energies become exact when $\nu\rightarrow\infty$ so should match the ED results in that limit.
\begin{figure}
\includegraphics[width=0.8\columnwidth, angle=270]{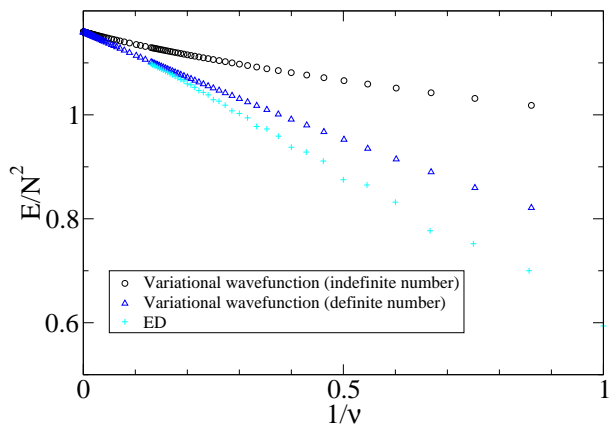}
\caption{Comparison of our results with ED data for $N_{\rm v}=6$ and $L_y/L_x=1/\sqrt{3}$. As expected and explained in section~\ref{variational} the indefinite particle number trial wavefunction overestimates the energy and this overestimate is larger at low filling factors. In all the other figures our variational results were obtained for the definite number trial wavefunction in Eq.~\ref{definite}}  \label{6vortices}
\end{figure}
\begin{figure}
\centering
\includegraphics[width=0.8\columnwidth, angle=270]{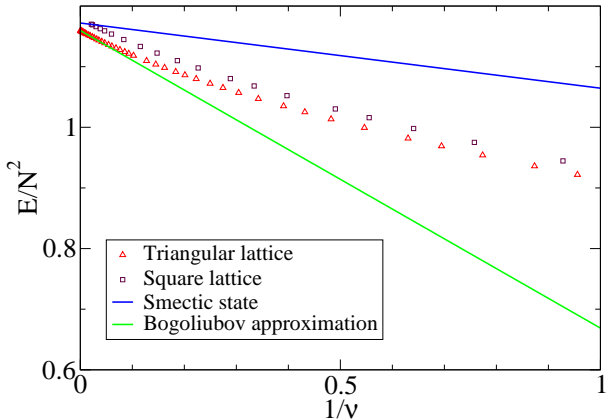}
\caption{The results for the largest studied system, i.e. for $N_{\rm v}=90$  ($L_y/L_x=\frac{5}{9}\sqrt{3}$ for the triangular lattice and $L_y/L_x=9/10$ for the square lattice). The variational energies were calculated using the definite-number trial wavefunction.} \label{90vortices}
\end{figure}
Fig.~\ref{6vortices} presents our variational bounds, obtained from both the indefinite-number and the definite-number trial wavefunctions, and compares them against the ED energies. Clearly, the state with definite particle number provides a much better variational bound. The energies of the definite-number state are accurate to within $1\%$, down to $\nu \sim 5$. (For filling factors $\nu\lesssim 5$ we expect
the groundstate to involve significant quantum fluctuations, being close to or in the regime of strong correlated quantum Hall states,
and the variational wavefunction for the vortex lattice to become inaccurate. Still, we present our results for the range $0\leq 1/\nu\leq 1$ for completeness.)

We can now take advantage of this success and study the behaviour of the condensate for large systems in the thermodynamic limit, well outside the range of the available ED data. 

Fig.~\ref{90vortices} shows the results of variational
  calculations for the largest system we have studied, of $N_{\rm
    v}=90$ vortices. The lowest energy variational state is the
  triangular lattice (shown by triangular symbols in
  Fig.~\ref{90vortices}). At large filling factors, $1/\nu \to 0$, the
  variational results match with the Bogoliubov theory. However, there
  are significant deviations from the Bogoliubov result at low filling
  factors. For $\nu \simeq 10$ the deviation is as
  large as the difference in energy we find between triangular and
  square lattices, showing that these non-linearities are
  quantitatively important in determining the nature of the low-energy
  phase.

\begin{figure}
\centering
\includegraphics[width=0.8\columnwidth, angle=270]{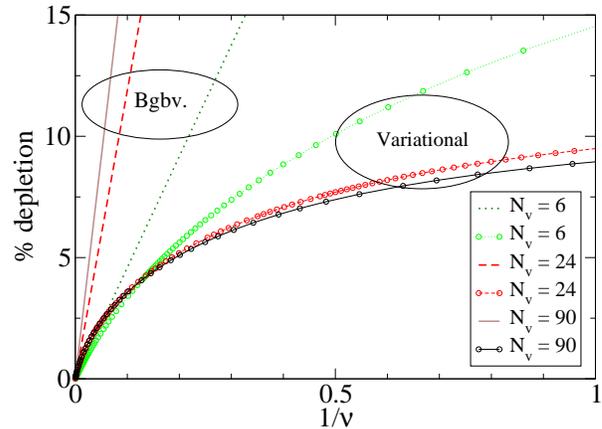}
\caption{Condensate depletion in a range of system sizes. The symbols correspond to our variational results for the definite-number state and the solid lines show the depletion calculated in the Bogoliubov approximation.} \label{depletion}
\end{figure}

We have also computed the condensate depletion, in a range of
  system sizes, as a function of the filling factor $\nu$ (see
Fig.~\ref{depletion}). The inclusion of the quartic term quenches
condensate depletion significantly. Already, at $\nu\sim10$, we see a
$15\%$ difference in condensate depletion for $N_{\rm v}=90$ when it is
included. The condensate depletion, although substantially
  reduced from the prediction of Bogoliubov theory, is still
  significant. An experimental observation of this reduction from
  $100\%$ condensate fraction -- even in the limit of vanishing
  temperature -- would be a clear indication of the effects of quantum
  fluctuations~\cite{shm2}.

\begin{figure}
\centering
\includegraphics[width=0.8\columnwidth, angle=270]{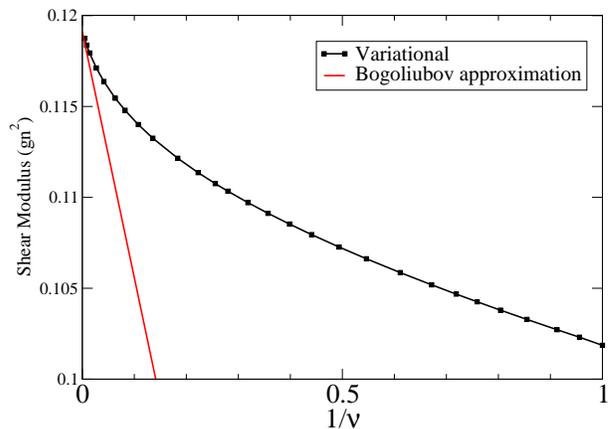}
\caption{Shear modulus ($C_2$) of the triangular vortex lattice as a function of the filling factor for $N_{\rm v}=90$} \label{shearmod90}
\end{figure}

As discussed in Sec~\ref{Bgbv}, condensate depletion leads to a non-vanishing vortex core density. Fig.~\ref{core} shows the vortex core density for $N_{\rm v}=90$, obtained from the variational estimates of the occupation numbers of $\mathbf{k \neq 0}$ states. The contribution of the quartic term is very significant and cannot be neglected at filling factors of the order of $N_{\rm v}\sim 90$, where Bogoliubov is no longer a valid approximation (see the scaling argument in Section.~\ref{variational}). 

It is particularly striking that, for filling factors as small
  as $\nu = 1$, the core density does not reach the critical value
$n_{\rm c}$ obtained from the Lindemann melting criterion with
  parameter $\langle\delta\mathbf{r}^2\rangle= 0.145l^2$. The
  application of this numerical value for the Lindemann criterion is
  highly questionable in this case of {\it quantum} melting of a
  vortex lattice.  Since it is known from numerical exact
  diagonalization studies that the lattice melts with $\nu_{\rm
    c}\simeq 2-6$~\cite{QuantumCooper,CooperSize,vedral}, our results
  for the core density suggest that the Lindemann parameter should be
  significantly reduced. We do not believe it is helpful to quote a
  number here, since the variational approach (based on expanding
  around the triangular vortex lattice) may be inaccurate in the
  vicinity of the melting transition.  Nevertheless, note that there
  is already a very clear departure from the Bogoliubov results for
  large $\nu \sim N_{\rm v}$. Our results show that the effects of
  quantum fluctuations in the positions of vortices are very much
  suppressed as compared to the expectations of Bogoliubov theory.

Including quantum fluctuations beyond the Bogoliubov approximation
also leads to very significant corrections to the shear modulus
of the vortex lattice.
Fig.~\ref{shearmod90} shows our numerical values of the
shear modulus for $N_{\rm v}=90$ and a range of filling factors, and
compares them against the Bogoliubov approximation. The 
shear modulus is not suppressed as much as 
predicted by Bogoliubov theory in Eq.~\ref{shear}.
The shear modulus can be experimentally measured from the frequencies of the Tkachenko sound modes. Following the continuum theory~\cite{Sonin, Sonin1} of the Tkachenko modes in a rapidly rotating BEC, the eigenfrequencies are given by
\begin{equation}
\omega_i = \sqrt{\frac{4\pi C_2}{\beta_{\triangle}g n^2\sqrt{3}}}\gamma_i(\Omega_r-\Omega),
\end{equation}
where $\gamma_1=7.17$ and $\gamma_2=16.9$ for the two lowest eigenmodes. The Tkachenko frequencies are proportional to $\sqrt{C_2}$. As described above, depending upon whether the oscillations are much faster or slower than the damping rates of Bogoliubov excitations, the system is in the collisionless or hydrodynamic regime. In the hydrodynamic regime zero-point fluctuations need to be taken into account. Our results show a $4\%$ shift from the mean-field value of $C_2$, at filling factors of $\sim 10$, giving an approximately $2\%$ shift in the Tkachenko eigenfrequencies. Observing this small softening of the shear modulus below its mean-field value would be an indication of the role of quantum fluctuations in the  vortex lattice phase.

\begin{figure}
\centering
\includegraphics[width=0.8\columnwidth, angle=270]{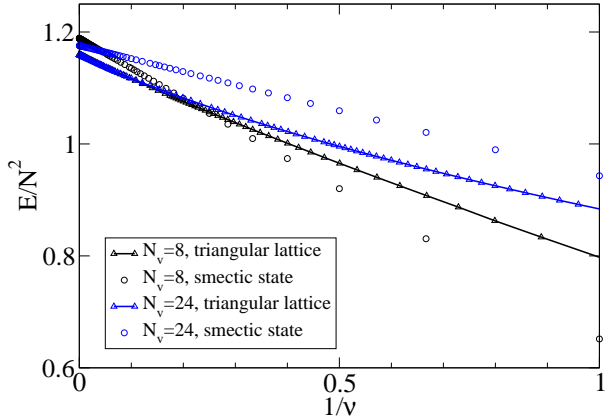}
\caption{Energy of the smectic state (circles) compared against the energy of the triangular vortex lattice (triangles) for $N_{\rm v}=8, L_y/L_x=\sqrt{3}/4$ and for $N_{\rm v}=24, L_y/L_x=4\sqrt{3}/9$} \label{smectic}
\end{figure}

\subsection{Competing Phases in Small Systems}

Finally, we turn to discuss the energetic stability of the triangular lattice phase compared with other competing phases in finite-sized systems. One possible phase the sytem might enter, as the filling factor or system size are reduced, is the smectic phase. In Ref.~\cite{CooperSize} the smectic state was shown to be favoured over the triangular lattice in small systems. The smectic state corrresponds to $N/N_{\rm s}$ independent condensates put into Landau-gauge states, which have the form of stripes aligned along the width of the system ($N_{\rm s}$ is the number of particles put into each state). We have compared our results for the vortex lattice phase with the energies of the stripe phase in finite-sized systems. Following Ref.~\cite{CooperSize}, the stripe separation that optimises the energy is given by $0.9083a_{\triangle}$. For each system we studied, we chose the separation that is closest to this value and commensurate with the length of the system, $L_x$. In the presence of a contact interaction, we only need to consider nearest-neighbour interactions between the stripes. This means that $\langle H_{\rm I}\rangle/N^2$ depends solely on the number of particles per stripe, $N_{\rm s}$, and the separation of the stripes, $\Delta x$. For $N_{\rm v}=4,6,8$, where $\Delta x=a_{\triangle}$ and there are two vortices per stripe ($N_{\rm v,s}=2$), the energy of the smectic state becomes
\begin{equation}
\langle H_{\rm I}\rangle/N^2=\frac{g}{2A}(1.1888-1.0746/N_{\rm s}),
\end{equation}
 for $N_{\rm v}=24$, where $\Delta x=\frac{\sqrt{3}}{2}a_{\triangle}$ and $N_{\rm v,s}=4$,
\begin{equation}
\langle H_{\rm I}\rangle/N^2=\frac{g}{2A}(1.1757-0.9307/N_{\rm s}),
\end{equation}
and for $N_{\rm v}=90$, where $\Delta x =\frac{9}{10}a_{\triangle}$ and $N_{\rm v,s}=9$,
\begin{equation}
\langle H_{\rm I}\rangle/N^2=\frac{g}{2A}(1.1720-0.9671/N_{\rm s}).
\end{equation}

Our results agree qualitatively with those of Ref.~\cite{CooperSize}.
In the thermodynamic limit the triangular vortex lattice has lower energy than the smectic state down to small filling factors at which we expect to see fractional quantum Hall states~\cite{QuantumCooper} (FQHSs). We also find that the smectic phase is favoured for small systems and low filling factors, and that there is evidence for phase transitions between the smectic state and the triangular vortex lattice. However, there are important quantitative differences from the results of Ref.~\cite{CooperSize}. That work did not include quantum fluctuations, obtaining $\frac{g\beta_{\triangle}}{2A}(N^2-N)$ for the energy of the triangular vortex lattice. By including the Bogoliubov and quartic terms in our energy calculations, we revise the estimates of the transition points between the triangular and smectic states. For $N_{\rm v}=4,6,8$ ($N_{\rm v,s}=2$) our results show a phase transition at $\nu=3-4$. As expected, quantum fluctuations stabilise the triangular vortex lattice and our estimate is lower than $\nu=8-13$, obtained following Ref.~\cite{CooperSize}. Indeed, we find that once the system size is as large as $N_{\rm v}=24$ (see Fig.~\ref{smectic}), there is no transition to a smectic state for any filling factors $\nu > 1$.

Our studies  show that another  competing phase is the square vortex lattice. Like the smectic state, it can be favoured over the triangular lattice in small systems and at low filling factors. Following the same variational method as for the triangular lattice, we computed the matrix elements in Eq.~\ref{matrix}, with the lattice vectors now describing a square lattice, and calculated variational energies using the definite-number state (Eq.~\ref{definite}). We found an energy cross-over between the two vortex lattice phases for $N_{\rm v}=6$ at $\nu\simeq 4$. In the largest, approximately isotropic, system that we studied, i.e. $N_{\rm v}=90$ (see Fig.~\ref{90vortices}), the triangular vortex lattice is lower in energy down to $\nu\simeq 1$.
Thus, our results show that, while quantum fluctuations reduce the energy gap between the square and triangular vortex lattice phases (and, as described above, 
also reduce the shear modulus of the triangular lattice),
the triangular vortex lattice remains the lower energy state in the thermodynamic limit.

\section{Summary}\label{Summary}

We have developed a variational method to describe the quantum
  fluctuations of 2D vortex lattice phases of ultracold atomic gases
  in the lowest Landau level regime.
   We find very good agreement (within $1\%$)
  between the results of our variational method and those of exact
  diagonalization studies at filling factors of 5 and above.
 Our theory includes the
  effects of interactions between the quantum fluctuations beyond the
Bogoliubov description. These are found to have significant
quantitative  effects
on physical properties.
 Our theory predicts
dramatically reduced condensate depletion as compared to Bogoliubov
theory, and dramatic reduction in the extent to which the positions of the vortices fluctuate, as evidenced by the quantum contribution to the particle density at the vortex cores. Measurements of the condensate fraction,  of the shear modulus, or of the particle density at the vortex cores could provide signatures of the quantum fluctuations of vortices. Our results provide clear predictions for the sizes of these effects, which differ markedly from the predictions of previous theories even at relatively large filling factors. Finally, we used our theory
to assess the competition of the triangular vortex lattice phase with
the smectic and square phases in finite-size systems.

\acknowledgements{We gratefully acknowledge financial support from EPSRC.}

\appendix

\section{} \label{Appendix}
Here we outline how to express the matrix elements in terms of analytic functions. We can sum over the first pair of variables in Eq.~\ref{matrix}, say $m_{11}$ and $m_{12}$, using the identity~\cite{proof}
\begin{eqnarray}
&&\sum_{m,n}(-1)^{mn}e^{-\pi t (m^2+n^2) -\pi t mn+2mix+2niy}\nonumber\\
&&\equiv\vartheta_3(x|it)\vartheta_3(2y-x|3it)-
\vartheta_1(x|it)\vartheta_1(2y-x|3it),\nonumber\\ \label{Identity}
\end{eqnarray}
where $\vartheta_i$ are the Jacobi Theta-functions and $t=1/\sqrt{3}$. However, $x$ and $y$ will themselves be dependent on other summation variables ($m_{31}$ and $m_{32}$ in this case), which need to be extracted out of the Theta-function arguments before they too can be summed over using Eq.~\ref{Identity}. They can only be extracted if their coefficients are multiples of $\pi/2$ or $it/2$, where $it$ is the period of a given Theta-Function. A number of identities that exploit the quasi-periodicity of the Theta-functions were used in this 'extraction' (see Ref.~\cite{Whittaker} for an exhaustive list).

\section{}
Here we derive the relation between the vortex core density and the mean-squared vortex core displacement. We take the ansatz that vortex core positions are described by the following Gaussian probability distribution:
\begin{equation}
p(\delta\mathbf{r})=\frac{1}{\pi \langle \delta\mathbf{r}^2\rangle} e^{-\delta \mathbf{r}^2/\langle \delta\mathbf{r}^2\rangle},
\end{equation}
where $\delta\mathbf{r}$ is the displacement of the vortex core from its mean-field position and $\langle \delta\mathbf{r}^2\rangle=\xi l^2$ is the variance of that displacement. It turns out, however, that the relation we are deriving is independent of the above ansatz when the vortex core fluctuations are small $|\delta\mathbf{r}| \ll l$, which is the case right up to the melting point given by the Lindemann criterion that we use ($\xi \sim 0.145$). The average particle density at the vortex core is then given by
\begin{eqnarray}
N\langle|\Psi_{\mathbf{0}}(\mathbf{r_{\rm{core}}})|^2\rangle&=&\frac{2 n}{(\xi+2)\zeta(\mathbf{0})}\sum_{\mathbf{m,n}}(-1)^{n_1 n_2}e^{-\mathbf{r_n}^2/4 l^2} \nonumber\\
&&\times e^{-(\mathbf{r_m}-\frac{\mathbf{a_1}}{2}-\frac{\mathbf{a_2}}{2})^2/ \xi l^2}  \nonumber \\
&&\times e^{\frac{1}{2 l^2 (\xi +2)}\mathbf{r_n\cdot(a_1+a_2-2 r_m)}} \nonumber\\
&&\times e^{-\frac{i}{2 l^2 (\xi +2)}\hat{\mathbf{z}}\cdot(\mathbf{r_n\times (a_1+a_2-2 r_m)})}  \nonumber\\
&&\times e^{\frac{1}{2 l^2 \xi(\xi+2)}(\mathbf{a_1+a_2-2 r_m)^2}},
\end{eqnarray}
where $n=N/A$ is the mean particle density.

\section{}

We obtain the expectation value of $H_{\rm I}$ for the trial wavefunction given in Eq.~\ref{definite}. Let 
\begin{eqnarray}
U_N=\exp\left[-\sum_{\mathbf{k}\neq\mathbf{0}} \frac{\theta_\mathbf{k}}{2}\left(\beta b^{\dagger}_{\mathbf{k}}b^{\dagger}_{-\mathbf{k}}e^{i\phi_\mathbf{k}}-\beta^{-1}b_{\mathbf{k}}b_{-\mathbf{k}}e^{-i\phi_\mathbf{k}}\right)\right], \nonumber \\
\end{eqnarray}
and $|\Psi_{\rm{trial}}\rangle=U_N|N\rangle$, where $U_N$ is a depletion operator that conserves total particle number.
We have made use of the following identities which, with the exception of the last one, mirror those given in Eq.~\ref{identity}:
\begin{eqnarray}
&&U_N^{\dagger}b_\mathbf{q}U_N\equiv b_\mathbf{q}\cosh\theta_\mathbf{q} -b^{\dagger}_{-\mathbf{q}}\beta e^{i\phi_\mathbf{q}}\sinh\theta_\mathbf{q}, \nonumber \\
&&U_N^{\dagger}b^{\dagger}_{\mathbf{q}}U_N\equiv b^{\dagger}_{\mathbf{q}}\cosh\theta_\mathbf{q} -b_{-\mathbf{q}} \beta^{-1}e^{-i\phi_\mathbf{q}}\sinh\theta_\mathbf{q}, \nonumber \\
&&U_N^{\dagger}b_0b_0U_N=\beta\sqrt{N-1}\sqrt{N}\left[1-\frac{2N-1}{2N(N-1)}U_N^{\dagger}D U_N\right]\nonumber\\
&&+ \mathcal{O} \left(D^2/N\right), 
\end{eqnarray}
where $D=\sum_{\mathbf{q}\neq\mathbf{0}} b^{\dagger}_{\mathbf{q}}b_\mathbf{q}$ is the depletion operator. Unlike $U$, $U_N$ does not commute with $b_0$ and $b_0^{\dagger}$, and therefore a third identity was required.      After a somewhat lengthy calculation one finds

\begin{widetext}
\begin{eqnarray}
&&\langle \Psi_{\rm{trial}}|H_{\rm I}|\Psi_{\rm{trial}}\rangle =M(\mathbf{0,0,0,0})\left[N^2-N+(1-2N)\sum_{\mathbf{q}\neq\mathbf{0}}\sinh^2\theta_\mathbf{q} +\left(\sum_{\mathbf{q}\neq\mathbf{0}}\sinh^2\theta_\mathbf{q} \right)^2 +2\sum_{\mathbf{q}\neq\mathbf{0}}\sinh^2\theta_\mathbf{q} \cosh^2\theta_\mathbf{q}    \right] \nonumber \\
&&+4\sum_{\mathbf{q}\neq\mathbf{0}}M(\mathbf{0,q,0,q}) \sinh^2\theta_\mathbf{q} \left(N -\sum_{\mathbf{k}\neq\mathbf{0}}\sinh^2\theta_\mathbf{k} - 2\cosh^2\theta_\mathbf{q} \right) + \sum_{\mathbf{k}\neq \mathbf{0},\mathbf{q}\neq\mathbf{0}}2\sinh^2\theta_\mathbf{k}\sinh^2\theta_\mathbf{q} M(\mathbf{k,0,-k+q,-k+q})\nonumber \\
&&+\frac{1}{2}\sqrt{N^2-N}\sum_{\mathbf{q}\neq\mathbf{0}}\sinh2\theta_\mathbf{q} \left[\frac{2N-1}{N(N-1)}\left(\sinh^2\theta_\mathbf{q}+\frac{1}{2}\sum_{\mathbf{k}\neq\mathbf{0}}\sinh^2\theta_\mathbf{k}\right)-1\right]\left(e^{i\phi_\mathbf{q}}M(\mathbf{q,-q,-q},-2\mathbf{q})+ e^{-i\phi_\mathbf{q}}M(\mathbf{0,q,-q,0})\right) \nonumber\\
&&+\sum_{\mathbf{k}\neq \mathbf{0},\mathbf{q}\neq\mathbf{0}}\sinh\theta_\mathbf{q}\cosh\theta_\mathbf{q}\sinh\theta_\mathbf{k}\cosh\theta_\mathbf{k} e^{i(\phi_\mathbf{k}-\phi_\mathbf{q})}M(\mathbf{k,-k+q,-q-k,-2k})+\mathcal{O}\left[\frac{1}{N}\left(\sum_{\mathbf{q}\neq\mathbf{0}}\sinh^2\theta_\mathbf{q} \right)^2\right].\nonumber\\
\end{eqnarray}
\end{widetext}

When finding the set of parameters $\theta_\mathbf{k}$ and $\phi_\mathbf{k}$ that minimises the expectation value of $H_{\rm I}$, we need to check that the error $\sim \mathcal{O}\left((N-N_0)^2/N\right)$ is small and can be neglected. In other words, we are constrained to optimise in the region of parameter space where the incurred error is small.


\begin{thebibliography}{99}
\bibitem{blochdz} I. Bloch, J. Dalibard, and W. Zwerger, Rev. Mod. Phys. \textbf{80}, 885, (2008).
\bibitem{Donnelly} R. J. Donnelly, \textit{Quantized Vortices in Helium II}, Cambridge University Press, 1991.
\bibitem{fetterreview} A. L. Fetter, Rev. Mod. Phys. \textbf{81}, 647, (2009).
\bibitem{madison} K. W. Madison, F. Chevy, W. Wohlleben, and J. Dalibard, Phys. Rev. Lett. \textbf{84}, 806, (2000).
\bibitem{aboshaeer} J. R. Abo-Shaeer, C. Raman, J. M. Vogels, and W. Ketterle, Science \textbf{292}, 476, (2001).
\bibitem{SchweikhardCEMC92} V. Schweikhard, I. Coddington, P. Engels, V. P. Mogendorff, E. A. Cornell, Phys. Rev. Lett. \textbf{92}, 040404, (2004).
\bibitem{cornell1} I. Coddington, P. Engels, V. Schweikhard, and E. A. Cornell, Phys. Rev. Lett. \textbf{91}, 100402, (2003).
\bibitem{foot} N. L. Smith, W. H. Heathcote, J. M. Krueger, and C. J. Foot, Phys. Rev. Lett. \textbf{93}, 080406, (2004).
\bibitem{CooperReview} N. R. Cooper, Adv. Phys. \textbf{57}, 539 (2008).
\bibitem{QuantumCooper} N. R. Cooper, N. K. Wilkin, and J. M. F. Gunn, Phys. Rev. Lett. \textbf{87}, 120405 (2001).
\bibitem{CooperSize} N. R. Cooper and E. H. Rezayi, Phys. Rev. A \textbf{75}, 013627 (2007).
\bibitem{vedral} Z. Liu, H. Guo, V. Vedral, H. Fan, Phys. Rev. A \textbf{83}, 013620, (2011).
\bibitem{Sinova} J. Sinova, C. B. Hanna, and A. H. MacDonald, Phys. Rev. Lett. \textbf{89}, 030403 (2002).
\bibitem{Baym} G. Baym, Phys. Rev. A \textbf{69}, 043618, (2004).
\bibitem{Sonin} E. B. Sonin, Phys. Rev. A \textbf{72}, 021606(R) (2005).
\bibitem{Matveenko} S. I. Matveenko and G. V. Shlyapnikov, Phys. Rev. A \textbf{83}, 033604 (2011).
\bibitem{AxialConfinement} Axially, the energy spectrum is quantised as follows: $E_m=\hbar\Omega_z(m+\frac{1}{2})$. To a good approximation, the single particle states are restricted to $m=0$ i.e. to quasi-2D.
\bibitem{bournewilkingunn} A. Bourne, N. K. Wilkin, J. M. F. Gunn, Phys. Rev. Lett. \textbf{96}, 240401 (2006).
\bibitem{Rashba} E. I. Rashba, L. E. Zhukov, and A. L. Efros, Phys. Rev. B \textbf{55}, 5306 (1997).
\bibitem{Zak} J. Zak, Phys. Rev. \textbf{134}, A1602 (1964).
\bibitem{Burkov} A. A. Burkov, Phys. Rev. B \textbf{81}, 125111 (2010).
\bibitem{SinovaLind} A. Rozhkov and D. Stroud, Phys. Rev. B \textbf{54}, R12697 (1996).
\bibitem{Bosons} M. Girardeau, R. Arnowitt, Phys. Rev. \textbf{113}, 755 (1959). 
\bibitem{shm2} J. Sinova, C. B. Hanna, A. H. MacDonald, Phys. Rev. Lett. \textbf{90}, 120401, (2003).
\bibitem{Sonin1} E. B. Sonin, Phys. Rev. A \textbf{71} 011603(R) (2005).
\bibitem{proof} This can be proven by summing over $n$ first and then applying the Jacobi imaginary transformation. See, for example, \textit{Higher Transcendental Functions}, edited by A. Erd\'{e}lyi (McGraw-Hill, New York, 1953), Vol. II, Chap. XIII.
\bibitem{Whittaker} E. T. Whittaker and G. N. Watson, \textit{A Course in Modern Analysis}, fourth edition, Cambridge University Press, 1927, Chap. XXI.


\end{thebibliography}
\end{document}